\documentclass[aps,prl,reprint,amsmath,amssymb,showpacs,nobibnotes]{revtex4-1}
\usepackage{graphicx}
\usepackage{hyperref} 
\usepackage{dsfont}    
\usepackage{color}

\newcommand{\rrr}[1]{\vskip 0.2cm \noindent{\it #1} ---}

\begin{document}
\title{Bound of Casimir Effect by Holography}
\author{Rong-Xin Miao}
\affiliation{School of Physics and Astronomy, Sun Yat-Sen University, Zhuhai, 519082, China}


\begin{abstract}
Inspired by the Kovtun-Son-Starinet bound, we propose that holography imposes a lower bound on the Casimir effect. For simplicity, we focus on the Casimir effect between parallel planes for three-dimensional conformal field theories and briefly comment on the generalizations to other boundary shapes and higher dimensions. Remarkably, the ghost-free holographic models impose a universal lower bound on the Casimir effect. We verify the holographic bound by free theories, the Ising model, and $O(N)$ models with $N=2,3$ at critical points and prove it for the two-dimensional case. Remarkably, a general class of quantum field theories without conformal symmetries also obeys the holographic bound. 
\end{abstract}

\maketitle


\rrr{Introduction} Casimir effect \cite{Casimir:1948dh} is an interesting quantum effect that originates from the changes of vacuum fluctuations due to the boundary \cite{Plunien:1986ca, Bordag:2001qi, Milton:2004ya, Bordag:2009zz}. It is a candidate of dark energy \cite{Wang:2016och} and plays a vital role in QCD  \cite{Vepstas:1984sw} and wormhole physics \cite{Morris:1988tu, Maldacena:2020sxe}. It has been measured in experiments \cite{Mohideen:1998iz, Bressi:2002fr, Klimchitskaya:2009cw} and has significant potential applications in nanotechnology. As the vacuum energy, Casimir energy is defined as the lowest energy of a quantum system with a boundary. Since energy should be bounded from below, it is expected that there is a fundamental lower bound of the Casimir effect. This letter explores this vital problem. Holography (AdS/CFT) \cite{Maldacena:1997re} provides a powerful tool to study the strongly coupled conformal field theory (CFT) and usually imposes bounds on various physical phenomena. The most famous example is the Kovtun-Son-Starinet (KSS) bound \cite{Kovtun:2004de} for hydromechanics, which conjectures that the ratio of shear viscosity $\eta$ to entropy density $s$ has a lower bound 
\begin{eqnarray}\label{KSS bound} 
\frac{\eta}{s}\ge \frac{\hbar}{4\pi k_B},
\end{eqnarray}
where $\hbar$ is the reduced Planck constant and $k_B$ is the Boltzmann constant. The lower bound is obtained from the strongly coupled CFT dual to Einstein gravity \cite{Policastro:2001yc} and can be modified slightly by higher derivative gravity \cite{Brigante:2007nu, Kats:2007mq, Brigante:2008gz}. 

Now, let us explore whether AdS/CFT sets a similar lower bound for the Casimir effect. For simplicity, we focus on the Casimir effect for three-dimensional (3D) CFTs between parallel planes (strip) and comment on various generalizations at the end of the Letter. We consider flat space and impose the same boundary conditions on the two planes.
Then, the Casimir effect takes a universal form \cite{Miao:2024ddp}
\begin{eqnarray}\label{Tij strip} 
\langle T^i_{\ j} \rangle_{\text{strip}}=\kappa_1 \frac{ \hbar c}{L^3}\text{diag}\Big(1, -2,1\Big),
\end{eqnarray}
where $\kappa_1$ is the dimensionless Casimir amplitude, $c$ is the speed of light, and $L$ is the strip width. For the fixed $L$, $-\kappa_1$ should be bounded from below to avoid negative infinity energy density $T_{tt}=-\kappa_1 \hbar c/L^3$ and pressure $T_{nn}=-2\kappa_1 \hbar c/L^3$. Note that the Casimir amplitude $\kappa_1$ is an extensive parameter in the sense that if one CFT has a Casimir amplitude $\kappa_1$, then $N$-copies of the same CFT have a Casimir amplitude $  N\kappa_1$. It implies that only $-\kappa_1$ over another extensive parameter could have a lower bound. Natural candidates of these extensive parameters are boundary central charges in Weyl anomaly\cite{Jensen:2015swa} $\mathcal{A}=\int_{\partial M} d^2x \sqrt{|\sigma|} \Big(a R_{\partial M} +b \text{tr}\bar{k}^2\Big)$, where $R_{\partial M} $ and $\bar{k}$ are the intrinsic Ricci scalar and traceless parts of extrinsic curvatures on the boundary $\partial M$, respectively. $a$ is the A-type boundary central charge, which decreases under the boundary renomalization group (RG) flow \cite{Jensen:2015swa}. We remark that $a$ can be positive, zero, or negative. For example, $a=\pm 1/(384 \pi)$ for conformally coupled free scalar with Robin boundary condition (RBC) and Dirichlet boundary condition (DBC) \cite{Jensen:2015swa}, respectively; $a\sim \sinh(\rho)$ with $-\infty<\rho<\infty$ for holographic CFT$_3$ \cite{Fujita:2011fp}. Since $a$ can be zero and negative, there is no lower bound for $-\kappa_1/a$. On the other hand, the B-type boundary central charge $b$ is always positive since it is related to the Zamolodchikov norm of the displacement operator, i.e.,  $ b=C_D\pi/32>0$. The displacement operator $D(y)$ describes the violation of space translation invariance normal to the boundary \cite{Billo:2016cpy}, i.e., $\nabla_i T^{ij}=\delta(x)D(y) n^j$. The two-point function defines its Zamolodchikov norm, i.e., 
$\langle D(y) D(0) \rangle =C_D c^2 \hbar^2/|y|^{6}$. By definition, $C_D$ is dimensionless and positive. 
The above arguments suggest $-\kappa_1/C_D$ should have a physical lower bound. Usually, a bound is set either by free theories or strongly coupled CFTs dual to gravity. We have checked all the known examples and find that, similar to the KSS bound, holography gives the smallest $-\kappa_1/C_D$. Thus, we propose that the ratio of Casimir amplitude to the norm of the displacement operator has a lower bound set by holography
\begin{eqnarray}\label{3d Casimir bound}
-\frac{\kappa_1}{C_D} \ge -\frac{\pi ^{5/2} \Gamma \left(\frac{1}{3}\right)^3}{108 \Gamma \left(\frac{5}{6}\right)^3}\approx -2.17, \ \ \ \text{for 3D CFTs}.
\end{eqnarray}
It means the energy density $-\kappa_1/L^3$ per degree of freedom $C_D$ is bounded from below for a physical system with fixed size $L$. Remarkably, unlike the KSS bound (\ref{KSS bound}), different holographic models give the same universal lower bound (\ref{3d Casimir bound}) for the Casimir effect. We verify the lower bound (\ref{3d Casimir bound}) by free CFTs, the Ising model, and the $O(N)$ model with $N=2,3$ at critical points. Furthermore, we prove the corresponding bound for two-dimensional (2D) CFTs, which strongly supports our general proposal. Below, we discuss the test of our proposal, the derivation of the holographic bound, and its various generalizations. For simplicity, we focus on natural units with $c=\hbar=k_B=1$ in the following.

\begin{table}[ht]
\caption{$(-\kappa_1/C_D)$ for various 3D CFTs}
\begin{center}
    \begin{tabular}{| c | c | c | c |  c | c | c | c| c| c|c| }
    \hline
     Dirac fermion& Scalar & Ising & O(2)& O(3)&  AdS/BCFT \\ \hline
 $ -1.42 $ &  $ -1.89$ & -2.12 & -1.83 & -1.62 &$ -2.17$\\ \hline    \end{tabular}
\end{center}
\label{table1 Casimir bounds}
\end{table}

\rrr{Tests} 
Let us test the holographic bound (\ref{3d Casimir bound}) by various 3D CFTs, which is summarized in Table \ref{table1 Casimir bounds}. We have  $-\kappa_1/C_D=-3\pi  \zeta (3)/8\approx -1.42$ for Dirac fermions with bag boundary condition (BC) \cite{Bellucci:2009hh,McAvity:1993ue}, and $-\kappa_1/C_D=-\pi  \zeta (3)/2\approx -1.89$ for free scalars with RBC and DBC \cite{Romeo:2000wt,Miao:2018dvm}. For Ising model with $\phi^4$ interaction at the critical point, we have $2\kappa_1\approx 0.820$ and $C_D\approx  0.193$ \cite{Toldin:2021kun,Ising referee1,Ising referee2}, which yields $-\kappa_1/C_D\approx -2.12$. Here, we impose with $(++)$ BC for the Ising model, where the spins on the two boundaries have the same signs. Near the strip boundary, the scalar operator behaves as \cite{Cardy:1990xm}
\begin{eqnarray}\label{sect3: O}
\langle \sigma(z) \rangle=\frac{a_{\sigma}}{(2z)^{\Delta_{ \sigma} }}\Big(1+ B_{\sigma} (\frac{z}{L})^d+...\Big),
\end{eqnarray}
where $\Delta_{ \sigma}$ is the conformal dimension and $B_{\sigma}$ obeys the universal relation
\begin{eqnarray}\label{sect3: key relation}
\frac{B_{\sigma}}{\Delta_{ \sigma} } =2^{d-1} (d-1) \pi ^{-\frac{d}{2}} \Gamma \left(\frac{d}{2}\right) \ \frac{ \kappa_1}{C_D}.
\end{eqnarray}
Here, we use the notation of \cite{Toldin:2021kun} with $d=3$ for 3D CFTs.  For $O(N)$ model with $N=2,3$ and open BCs, \cite{Toldin:2021kun} obtains $B_{\sigma}\approx 1.21, 1.07$ and $\Delta_{ \sigma}\approx 0.519088, 0.518920$ by applying Monte Carlo simulations. Then (\ref{sect3: key relation}) gives $-\kappa_1/C_D\approx -1.83, -1.62$ for $N=2,3$, respectively. We summarize all the above results in the Table \ref{table1 Casimir bounds}, which all obey the holographic bound (\ref{3d Casimir bound}) will be derived below. It strongly supports our proposal. Interestingly, (\ref{sect3: key relation}) suggests the bound of the Casimir effect imposes an upper bound of $B_{\sigma}/\Delta_{ \sigma}$.

\rrr{Holographic bound} Let us study the holographic bound of the Casimir effect of a strip in AdS/BCFT \cite{Takayanagi:2011zk}. We are interested in the general ghost-free gravity in AdS$_4$, which turns out to be Dvali-Gabadadze-Porrati (DGP) gravity,\cite{Dvali:2000hr} 
\begin{eqnarray}\label{DGP gravity}
I=\int_N d^{4}x \sqrt{|g|} \Big(R +6 \Big)+2\int_Q d^3y \sqrt{|h|} \Big(K-T +\lambda \mathcal{R} \Big),
\end{eqnarray}
where $K$ is the extrinsic curvature, $T$ is the brane tension, $\lambda$ is the DGP parameter,  $R$ and $\mathcal{R}$ are Ricci scalars in bulk $N$ and on the brane $Q$, respectively. We have set Newton's constant $16\pi G_N=1$ and AdS radius $l=1$ for simplicity. Note that higher derivative gravity and negative DGP gravity generally suffer the ghost problems \cite{Miao:2023mui}, while Gauss-Bonnet/Lovelock gravity appears only in dimensions higher than four. Thus, positive DGP gravity with $\lambda\ge 0$ is the general ghost-free gravity theory in AdS$_4$. The case $\lambda=0$ corresponds to Einstein gravity. 
Following \cite{Takayanagi:2011zk}, we impose the Neumann boundary condition (NBC) on the brane $Q$
\begin{eqnarray}\label{sect2: NBC}
K^{ij}-(K-T+\lambda \mathcal{R}) h^{ij}+2 \lambda \mathcal{R}^{ij}=0,
\end{eqnarray}
where $T=2 \tanh(\rho)-2\lambda \text{sech}^2(\rho)$ is a useful parametrization \cite{Miao:2023mui}. The holographic norm of the displacement operator is given by \cite{Miao:2023mui, Miao:2018dvm}
\begin{eqnarray}\label{sect2: CD 3d}
C_D=\frac{32}{\pi  \left(\frac{2 \lambda }{2 \lambda  \sinh (\rho )+\cosh (\rho )}+2 \tan ^{-1}\left(\tanh \left(\frac{\rho }{2}\right)\right)+\frac{\pi }{2}\right)}. 
\end{eqnarray}
$C_D\ge 0$ together with $\lambda\ge 0$ yields the constraints: $0\le \lambda$ for $\rho \ge 0$; $0\le  \lambda \le -\frac{\coth (\rho )}{2}\ge 1/2$ for $\rho \le 0$. Below, we will show that $\lambda=1/2$ is a phase-transition point and there are other physical constraints for $\lambda$. In the following, we show only key points and leave the detailed derivations to the supplemental material. 

\begin{figure}[t]
\centering
\includegraphics[width=6.9cm]{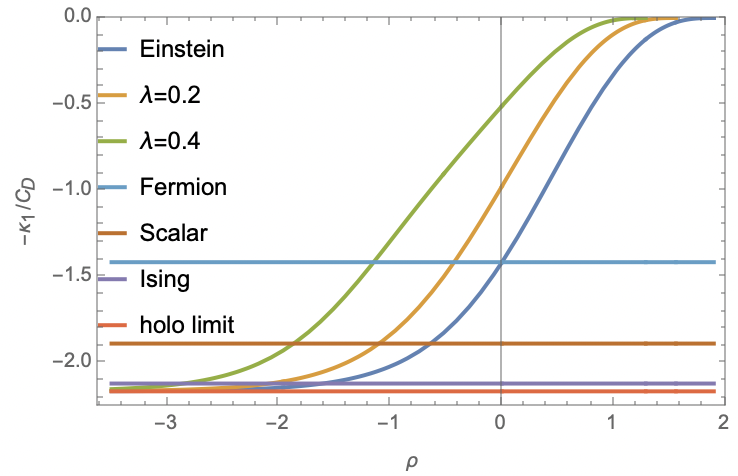}  \includegraphics[width=6.9cm]{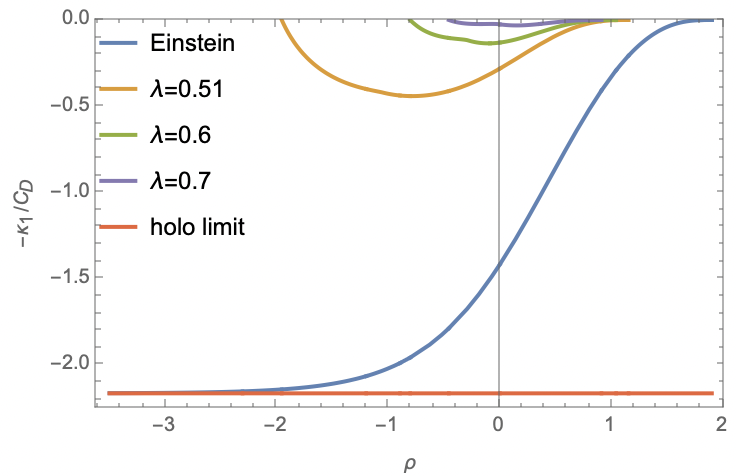}
\caption{$-\kappa_1/C_D$ in normal phase with $0\le \lambda< 1/2$ (top) and singular phase with $1/2< \lambda\le1$ (bottom) for DGP gravity. Einstein gravity corresponds to $\lambda=0$. The top figure shows all holographic curves approach the same limit (red line) from above for $\rho\to -\infty$ ($T\to -2$) in the normal phase.  The bottom figure shows $-\kappa_1/C_D$ increases with $\lambda$ and approaches zero for $\lambda\to 1$ in the singular phase. Together, it shows all examples obey the holographic bound eq.(\ref{3d Casimir bound}) (red line).}
\label{bound1}
\end{figure}

The vacuum of a strip is dual to the AdS soliton \cite{Fujita:2011fp}
\begin{eqnarray}\label{sect2: AdS soliton}
ds^2=\frac{\frac{dz^2}{h(z)}+h(z)d\theta^2-dt^2+dy^2}{z^2},
\end{eqnarray}
where $h(z)=1-z^3/z_h^3$. The absence of conical singularities fixes the period of angle $\theta$ in bulk $\beta=4\pi/|h'(z_h)|=4z_h\pi/3$. The strip is defined on the AdS boundary $z=0$ with $0\le \theta \le L$, where $L\ge 0$ is the strip width. We derive the energy density $T_{tt}=-1/z_h^3$ from (\ref{sect2: AdS soliton}) by applying the holographic renormalization \cite{deHaro:2000vlm}. Compared with the Casimir effect (\ref{Tij strip}), we get the holographic Casimir amplitude
\begin{eqnarray}\label{sect2: kappa1}
\kappa_1=L^3/z_h^3,
\end{eqnarray}
where $L$ can be fixed by NBC (\ref{sect2: NBC}). By using the rescale invariance of AdS, we set $z_h=1$ below. For negative brane tension $T\le 0$, 
we derive the strip width
\begin{eqnarray}\label{sect2: L I}
L_{\text{I}}=\int_0^{z_{\max}} \frac{2\Big(\sqrt{h\left(z_{\max }\right)}-\lambda  h(z)\Big)\  dz}{h(z) \sqrt{H(z, z_{\max })-h(z_{\max })}},
\end{eqnarray}
where $H=\frac{1}{2} h(z) \left(1+2 \lambda  \sqrt{h\left(z_{\max }\right)}+X\right)$,  $X=\sqrt{1+4 \lambda ^2 h(z)-4 \lambda  \sqrt{h\left(z_{\max }\right)}}$ and $z_{\text{max}}$ is the turning point obeying $T=-2 \sqrt{h(z_{\text{max}})}$. The strip width with $T>0$ can be obtained from the complement of that with $T<0$ by $L_{\text{II}}=\beta - L_{\text{I}}\Big(T\to -T, \lambda\to -\lambda\Big)$. It is the typical trick used in \cite{Fujita:2011fp}, based on extrinsic curvature flip signs when crossing the brane. In the viewpoint of NBC (\ref{sect2: NBC}), $T, \lambda$ change signs equivalently. Some comments are in order. First, $L_{\text{I}}$ (\ref{sect2: L I}) is negative for sufficiently large $\lambda$. The physical constraint $L\ge 0$ yields $\lambda\le 1$. Second, the case $0 \le \lambda< 1/2$ can continuously transform into that of Einstein gravity, while the case $1/2 < \lambda\le 1$ cannot. See Fig. \ref{bound1} for instance. We name them normal and singular phases, respectively. Third, the strip width $L$ decreases with brane tension $T$ \cite{Fujita:2011fp} and $L\ge 0$ sets an upper bound of $T$ \cite{Miyaji:2021ktr}. For $T\ge 0$, $L$ should be smaller than the angle period $\beta$ to avoid conical singularity. Luckily, $L\le \beta$ is satisfied automatically for $T\ge 0$ \cite{Miyaji:2021ktr}. On the other hand, there is no upper bound of $L$ for $T<0$, since the conical singularity is hidden behind the brane \cite{Tadashi}. Thus, it is irrelevant to the bulk dual of the strip. As a result, there is no extra constraint on the lower bound of $T$, and it can take the minimal value $T=-2$ \cite{Miyaji:2021ktr, Tadashi}. We remark that $T\ge -2$ sets no constraint on $\rho$ in the normal phase, while it sets a lower bound of $\rho$ in the singular phase. See Fig. \ref{bound1} (bottom). 

We can derive the ratio $-\kappa_1/C_D$ from (\ref{sect2: CD 3d},\ref{sect2: kappa1},\ref{sect2: L I}). Refer to Fig. \ref{bound1} for details. Fig. \ref{bound1} (top) denotes the normal phase with $0\le \lambda<1/2$, which shows that the smaller $\rho$, the smaller the $-\kappa_1/C_D$. Besides, all curves approach the same lower bound (red line) from above in the limit $T\to -2$. On the other hand, Fig. \ref{bound1} (bottom) labels the singular phase with $1/2\le \lambda\le 1$. It shows that $-\kappa_1/C_D$ with $\lambda>1/2$ is always larger than that of Einstein gravity (blue curve). Besides, the minimum of $-\kappa_1/C_D$ increases with $\lambda$ and approaches zero when $\lambda\to 1$. Since we are interested in the lower bound of ($-\kappa_1/C_D$), we focus on the normal phase below.

Let us explain the physical reasons why the minimal brane tension yields the lower bound of the Casimir effect in the normal phase. First, the Casimir energy equals the free energy at zero temperature, which can be calculated holographically by the Euclidean gravitational action. Naturally, the smaller the brane tension (energy), the smaller the Euclidean action, and, thus, the smaller the Casimir energy. Second, the brane tension decreases under boundary RG flow, i.e., $T_{\text{IR}}<T_{\text{UV}}$ \cite{Fujita:2011fp}.  At IR, many modes become inactive, resulting in reduced energy. Third, the Casimir effect is driven by long-range quantum correlations \cite{Dantchev:2022hvy}. As the tension (energy) decreases, these quantum correlations become stronger, which in turn increases the Casimir amplitude and reduces the Casimir energy. In summary, the minimal brane tension corresponds to the minimal Casimir energy.

Now, we give some analytical discussions. Performing coordinate transformation $z=z_{\text{max}} y$ and expanding (\ref{sect2: L I}) around $x=\text{sech}^2(\rho)\to 0$, we derive perturbatively for normal phase
\begin{eqnarray}\label{sect2.2: L I 2}
L_{\text{I}}= \frac{2 \sqrt{\pi } \Gamma \left(\frac{4}{3}\right)}{\Gamma \left(\frac{5}{6}\right)}\Big(\frac{(1-2 \lambda )^2}{x}\Big)^{\frac{1}{6}}+ O\left(\frac{x}{(1-2 \lambda )^2}\right)^{\frac{5}{6}}.
\end{eqnarray}
Taking limit $x\to 0$ ($T\to -2$) with $\lambda<1/2$, we obtain a universal limit independent of $\lambda$
\begin{eqnarray}\label{sect2.2: limit}
\lim_{T\to -2} \frac{-\kappa_1}{C_D}=\lim_{x\to 0} \frac{-L_{\text{I}}^3}{C_D}=-\frac{\pi ^{5/2} \Gamma \left(\frac{1}{3}\right)^3}{108 \Gamma \left(\frac{5}{6}\right)^3}.
\end{eqnarray}
It verifies our claim that different holographic models give the same lower bound (\ref{3d Casimir bound}) in the normal phase. The lower bound of $(-\kappa_1/C_D)$ is saturated by negative brane tension. We remark that the negative brane tension is well defined in AdS/BCFT \cite{Fujita:2011fp, Takayanagi:2011zk, Miyaji:2021ktr, Tadashi}. First, the brane tension is a cosmological constant rather than a kinetic-energy term on the end-of-the-world brane. Of course, the negative cosmological constant is well defined in AdS/CFT. Second, negative brane tension generally yields a negative A-type boundary central charge. But nothing goes wrong. Recall that the free scalar with the DBC has a negative A-type boundary central charge \cite{Jensen:2015swa}.

\rrr{General boundaries} Holography is expected to set a bound on the Casimir effect for general boundary shapes, not limited to strips. Let us provide some evidence. We first consider the wedge space, which is the simplest generalization of a strip. The Casimir effect takes the form
\begin{eqnarray}\label{Tij wedge} 
\langle T^i_{\ j} \rangle_{\text{wedge}}=\frac{f(\Omega) }{r^3}\text{diag}\Big(1, -2,1\Big),
\end{eqnarray}
where $f(\Omega)$ is the Casimir amplitude and $\Omega$ is the opening angle of wedge. In the limit $\Omega\to0, r\to \infty$ with $r\Omega=L$ fixed, the large $r$ region of the wedge approaches a strip. It results in $f(\Omega)\to \kappa_1/\Omega^3$ for $\Omega\to 0$  \cite{Deutsch:1978sc}. On the other hand, we have $f(\Omega)\to (C_D \pi /64) (\pi -\Omega)$ in the smooth limit $\Omega \to \pi$ \cite{Miao:2024ddp}. These two limits suggest AdS/BCFT with minimal brane tension sets the lower bound of the wedge Casimir effect
\begin{eqnarray}\label{wedge bound} 
\frac{-f(\Omega)}{C_D}\ge  \lim_{T\to -2}(\frac{-f(\Omega)}{C_D})_{\text{holo}}, \ \text{for}\  0<\Omega \le \pi.
\end{eqnarray}
See the following long paper \cite{Miao:2025utb} for the derivation and test of (\ref{wedge bound}). Let us go on to discuss the general boundary. Near any smooth boundary, the normal-normal component of the stress tensor takes a universal form in flat space \cite{Deutsch:1978sc, Miao:2017aba}
\begin{eqnarray}\label{general bdy} 
\langle T_{nn}(z) \rangle=\alpha \frac{\text{Tr}\bar{k}^2}{z}-\frac{\alpha}{2} k \text{Tr}\bar{k}^2 \log (\frac{z}{L})+ t_{nn}+ O(z) 
\end{eqnarray}
where $z$ is the distance to boundary, $L$ is the system size, and $t_{nn}$ denotes the finite term. The divergent terms appear since we consider the ideal boundary. There is a natural cutoff $z\ge \epsilon$ in real material with $\epsilon$ the lattice length. Remarkably, the displacement operator universally determines the divergent terms, i.e.,  $\alpha =-C_D \pi/16$ \cite{Miao:2017aba}. It strongly suggests $\langle T_{nn}(\epsilon) \rangle/C_D$, or equivalently, $ t_{nn}/C_D$ has a lower bound for fixed system size and boundary shape. A weaker proposal is for the average pressure 
\begin{eqnarray}\label{general bound} 
(\frac{\bar{t}_{nn}}{C_D})\ge  \lim_{T\to -2}(\frac{\bar{t}_{nn}}{C_D})_{\text{holo}},
\end{eqnarray}
where $\bar{t}_{nn}=(\int_{\partial M} t_{nn} dS)/ (\int_{\partial M} dS)$ is related to the variation of the total energy concerning the system size, i.e., $-dE/dL$.

\rrr{General dimensions} Let us comment on the holographic bound of the Casimir effect for a strip in general dimensions. The corresponding Casimir amplitude $\kappa_1$ and norm of displacement operator $C_D$ are given by \cite{Miao:2024ddp} for Einstein gravity.  By taking the limit $T\to -(d-1)$, we derive a universal lower bound,
\begin{eqnarray}\label{Casimir bound d}
\frac{-\kappa_1}{C_D}\ge\frac{-2^{d-2} d \pi ^{d-\frac{1}{2}}\Gamma \left(\frac{d-1}{2}\right) \Gamma \left(\frac{1}{d}\right)^d}{\Gamma (d+2) \left(d \Gamma \left(\frac{1}{2}+\frac{1}{d}\right)\right)^{d} }.
\end{eqnarray}
We verify that the free scalar obeys the bound, i.e., $(-\kappa_1/C_D)_{\text{scalar}}=-2^{1-d} \pi ^{d/2} \zeta (d)/\Gamma \left(\frac{d}{2}\right)\ge (\ref{Casimir bound d})$ \cite{Romeo:2000wt, Miao:2018dvm}. For $d=2$, we can prove the bound (\ref{Casimir bound d})
  \begin{eqnarray}\label{2d bound}
(\frac{-\kappa_1}{C_D})|_{d=2}\ge -\frac{\pi ^3}{12}.
 \end{eqnarray}
 According to \cite{Bloete:1986qm}, the inequality is saturated for general 2D CFTs with the same boundary conditions on the two planes. On the other hand, if we impose different boundary conditions on the two planes, we get larger Casimir energy \cite{Bachas:2006ti, Diatlyk:2024qpr}
   \begin{eqnarray}\label{mixed BC}
(-\kappa_1)_{\text{differ BC}} \ge (-\kappa_1)_{\text{same BC}}.
 \end{eqnarray}
Note that $C_D=c/(2\pi^2)$ is independent of the choices of boundary conditions, where $c$ is the central charge of 2d CFTs. Considering all the above statements, we get solid proof of the bound (\ref{2d bound}) for 2D CFTs, which is a strong support for our proposal (\ref{Casimir bound d}). 

\rrr{Non-CFTs} Interestingly, the lower bound of (\ref{3d Casimir bound}) also works for a general class of non-CFTs. Take an example of a 3D free scalar with mass $m$ and coupling $\xi$ with the Ricci scalar. It is a CFT only if $\xi=1/8$ and $m=0$. For non-CFT parameters, the energy density between parallel plates is no longer a constant \cite{Bordag:2009zz}. What is worse, it is divergent near the boundary. Fortunately, the total energy per area is finite \cite{Bordag:2009zz}. On the other hand, the pressure is still a finite constant. Thus, we define $\kappa_1=\langle T_{nn}(0) \rangle L^3/(-2)$ generally. By applying the textbook method, we obtain for both RBC and DBC, i.e., $16 \pi \kappa_1=-2 M^2 \log \left(1-e^{-2 M}\right)+2 M \text{Li}_2\left(e^{-2 M}\right)+\text{Li}_3\left(e^{-2 M}\right)$, where $M=m L>0$ is a dimensionless parameter, and $\text{Li}$ denotes the polylogarithmic function. Remarkably, $\kappa_1$ is independent of $\xi$ and decreases with mass,
\begin{eqnarray}\label{massive kappa1 derivative}
\frac{ d \kappa_1}{dM}=\frac{M^2 (1-\coth (M))}{8 \pi }<0.
\end{eqnarray}
For free non-CFTs, we can define $C_D$ via $\langle D(y) D(0)\rangle=C_D/|y|^6+O(1/|y|^5)$ in half-space, where $D=T_{nn}(0)$ \cite{Billo:2016cpy}. At a small distance $m |y|\ll 1$, the mass effect can be ignored. As a result, $C_D$ is independent of mass. The above discussions show $-\kappa_1/C_D$ is larger than that of a massless scalar, thus obeys the holographic lower bound (\ref{3d Casimir bound}). The above arguments apply to general massive free theories. The Casimir effect arises from long-range quantum correlations \cite{Dantchev:2022hvy}. When mass is introduced, these correlations become short range, which generally reduces the Casimir amplitude $\kappa_1$ \cite{Bordag:2009zz}. Recall that $C_D$ is mass independent. Consequently, $-\kappa_1/C_D$ reaches its minimum value in the massless limit for general free theories.

\rrr{Discussions} In conclusion, we propose that holography sets a fundamental bound on the ratio of the Casimir amplitude to the displacement operator norm. The Casimir amplitude is defined as the average pressure $\bar{t}_{nn}$ on the boundary. We take 3D CFTs in a strip as an example and then briefly discuss the generalization to other boundary shapes, higher dimensions, and free non-CFTs. It is interesting to test our proposal in experiments of critical systems \cite{Krech:Casimir Effect}. Finding a field-theoretical proof or counterexample for our proposal is also enjoyable. Our findings help estimate how much the Casimir effect contributes to dark energy and whether sufficient Casimir energy can be generated to support a traversable wormhole macroscopically. Additionally, it helps evaluate the maximum Casimir force between nanodevices, which can enhance nanotechnology designs.
Usually, a bound is set by either strongly coupled or free theories \cite{Kovtun:2004de, Hofman:2008ar, Hofman:2016awc}. 
For instance, the KSS bound \cite{Kovtun:2004de} for fluid is set by the strongly coupled theory dual to gravity, while the Hofman-Maldacena bound \cite{Hofman:2008ar, Hofman:2016awc} for bulk central charges is set by free theories. 
For entanglement entropy, there is a similar bound set by free theories, i.e., $(-F_0/C_T)\ge (-F_0/C_T)_{\text{free scalar}}$ \cite{Bueno:2023gey}, where $F_0$ is the universal coefficient of entanglement entropy and $C_T$ is a bulk central charge. We remark that the dual brane for entanglement entropy is tensionless while the one for the Casimir effect is tensive. With a tensive brane, we have more parameters to get a smaller lower bound of the Casimir effect in holography than in free theory. The Ising model of Table \ref{table1 Casimir bounds} also shows that a free scalar does not set the lower bound of the Casimir effect. In AdS$_3$, a conformal bootstrap restricts brane tension $|T|\le 0.99$ sightly stronger than $|T|\le 1$ \cite{Collier:2021ngi}. If this is also the case for both $T$ and DGP coupling $\lambda$ in higher dimensions, it may give a slightly stronger lower bound of the Casimir effect. Anyway, this Letter takes a significant step toward exploring the lower bound of the Casimir effect. 
Many open problems remain to be explored, particularly the generalization to non-CFTs with interactions.

 \section*{Acknowledgements}
We thank T.~Takayanagi and J. X. Lu for valuable comments and discussions. We are grateful to the conference ``Gauge Gravity Duality 2024,"where the work was completed. This work is supported by the National Natural Science Foundation of China (No.12275366). 

\appendix

\section{Bound of Casimir effect from DGP gravity}

This supplemental material studies the holographic bound of the Casimir effect in DGP gravity. We give detailed derivations of equations (11, 12, 13) in the main text. 

Let us quickly recall some key points. The vacuum of a strip is dual to the AdS soliton
\begin{eqnarray}\label{app a: AdS soliton}
ds^2=\frac{\frac{dz^2}{h(z)}+h(z)d\theta^2-dt^2+dy^2}{z^2},
\end{eqnarray}
where $h(z)=1-z^3/z_h^3$, and $\theta$ has the range $0\le \theta\le  L $ on the AdS boundary $z=0$. We label the embedding function of the brane $Q$ as
\begin{eqnarray}\label{app a: soliton Q}
\theta=S(z).
\end{eqnarray}
It obeys NBC 
 \begin{eqnarray}\label{app a: NBC}
K^{ij}-(K-T+\lambda \mathcal{R}) h^{ij}+2 \lambda \mathcal{R}^{ij}=0,
 \end{eqnarray} 
 where the tension is parameterized as
 \begin{eqnarray}\label{app a: Tension}
T=2 \tanh(\rho)-2\lambda \text{sech}^2(\rho). 
 \end{eqnarray} 
This parameterization can be derived from the gravity dual of a half-space. Since the region near any smooth boundary can be approximated by a half-space at the leading order of $z$, it works for general boundaries. Substituting (\ref{app a: soliton Q}) into NBC (\ref{app a: NBC}), we derive near the two boundaries of strip
  \begin{eqnarray}\label{app a: near strip bdy}
\theta=-\sinh(\rho) z+O(z^2), \  \ \theta-L=\sinh(\rho) z+O(z^2). 
 \end{eqnarray}  

\begin{figure}[t]
\centering
\includegraphics[width=7cm]{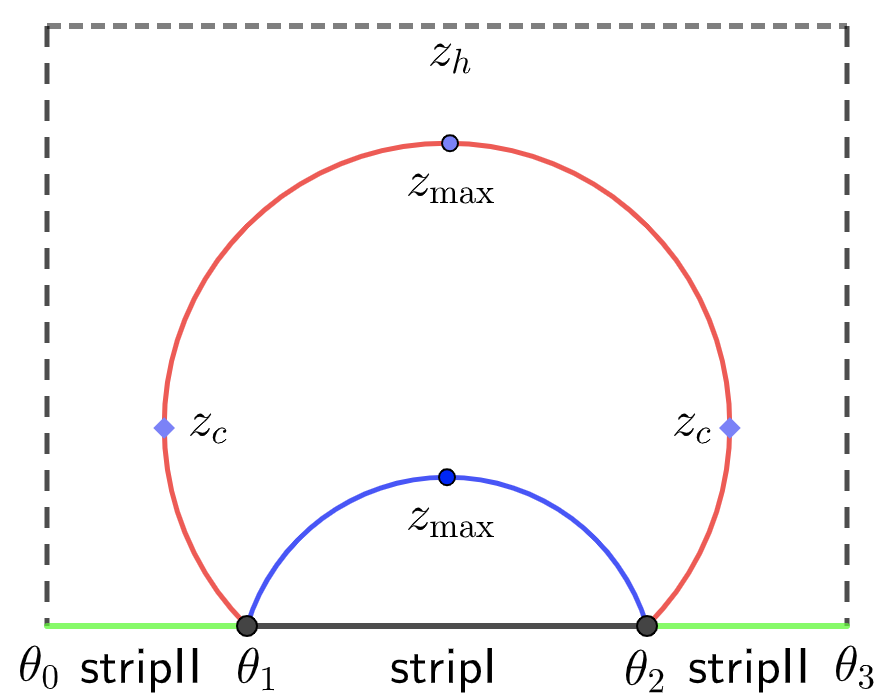}
\caption{Geometry of holographic strip: a portion of AdS soliton. The region between red/blue curves (branes) and black line is the bulk dual of strip I with negative brane tension $T\le 0$; its complement in bulk is the gravity dual for strip II (green line) with $T> 0$. The gravity dual of strip II contains the `horizon' $z=z_h$. To remove the conical singularity on it, the angle $\theta$ should be periodic. Thus green lines for strip II are connected. Without loss of generality, we focus on strip I with $T\le 0$. We have $(\rho<0)$ and $(\rho>0)$ for the blue and red curves, respectively. $z_{\text{max}}$ and $z_c$ denote the turning points with $\theta'(z_{\text{max}})=\infty$ and $\theta'(z_{c})=0$.} 
\label{holo strip}
\end{figure}

Let us first study the case with negative brane tension $T\le 0$; the other case with $T>0$ can be obtained by the complement of the first case. It is the typical trick used in \cite{Fujita:2011fp}. Let us explain more. See Fig. \ref{holo strip} for the geometry of the holographic dual of the strip. The region between red/blue curves (branes) and the black line is the bulk dual of strip I with $T\le 0$; its complement in the bulk is the gravity dual for strip II (green line) with $T> 0$. As shown in Fig. \ref{holo strip}, the gravity duals of strip I and strip II share the same EOW branes (blue or red curve depending on the theory parameter). As a result, the NBCs (\ref{app a: NBC}) for strip I and strip II take the same values. However, the extrinsic curvatures $K_{ij}$ flip signs, while the induced metric $h_{ij}$ and intrinsic Ricci tensor $\mathcal{R}_{ij}$ remain invariant when crossing the branes. From the viewpoint of NBC (\ref{app a: NBC}), $T$ and $\lambda$ equivalently change signs when crossing the branes. Note that the bulk dual for strip II contains the `horizon' $z=z_h$. To remove the conical singularity at $z=z_h$, we fix the period of angle $\theta$ as $\beta=4\pi/|h'(z_h)|=4\pi z_h/3$. Thus, for $T>0$, the left and right vertical dotted lines of  Fig. \ref{holo strip} are identified due to the periodicity of angle $\theta$. On the other hand, for strip I with $T\le 0$, the `horizon' and potential conical singularity is hidden behind the branes (red/blue curves). Thus, the conical singularity is irrelevant and $\theta$ can be non-periodic for $T\le 0$. 

Without loss of generality, we focus on the strip I with $T\le 0$ and $z_h=1$ below. According to  (\ref{app a: near strip bdy}), the blue and red curves of Fig. \ref{holo strip} correspond to $\rho<0$ and $\rho>0$, respectively. We focus on the left half of the branes; the right half can be obtained by symmetry. NBC (\ref{app a: NBC}) yields an independent equation for $T\le 0$
\begin{eqnarray}\label{app a: NBC solution}
T=-\frac{2 h(z) \left(h(z) S'(z) \sqrt{h(z) S'(z)^2+\frac{1}{h(z)}}+\lambda \right)}{h(z)^2 S'(z)^2+1}. 
\end{eqnarray}
Let us verify that it corresponds to the left half of the branes. From (\ref{app a: Tension},\ref{app a: NBC solution}), we derive $S'(0)=-\sinh(\rho)$ near $\theta_1$ of Fig. \ref{holo strip}, which agrees with the first equation of (\ref{app a: near strip bdy}). 
As shown in Fig. \ref{holo strip}, $S'(z_{\text{max}})=\infty$ and $S'(z_{c})=0$ at the turning points $z_{\text{max}}$ and $z_c$ for the left half of branes. Substituting $S'(z_{\text{max}})=\infty$ and $S'(z_{c})=0$ into (\ref{app a: NBC solution}), we obtain 
\begin{eqnarray}\label{app a: turning points}
T=-2 \sqrt{h(z_{\text{max}})}=-2\lambda h(z_c). 
\end{eqnarray}
Note that the turning point $z_c$ appears only in the red curve with $\rho>0$. Solving (\ref{app a: NBC solution}) and (\ref{app a: turning points}), we get 
\begin{eqnarray}\label{app a: dS}
S'(z)=\pm  \frac{\sqrt{h\left(z_{\max }\right)}-\lambda  h(z)}{h(z) \sqrt{H(z, z_{\max })-h(z_{\max })}},
\end{eqnarray}
where $H=\frac{1}{2} h(z) \left(1+2 \lambda  \sqrt{h\left(z_{\max }\right)}+X\right)$,  $X=\sqrt{1+4 \lambda ^2 h(z)-4 \lambda  \sqrt{h\left(z_{\max }\right)}}$. We choose the positive sign, i.e., $S'(z)\sim \Big(\sqrt{h\left(z_{\max }\right)}-\lambda  h(z)\Big) $, for the left halves of red and blue curves to get the correct behavior $S'(0)=-\sinh(\rho)$ near $\theta_1$ of Fig. \ref{holo strip}. The negative sign of (\ref{app a: dS}) corresponds to the right halves of branes. From (\ref{app a: soliton Q},\ref{app a: dS}), we derive the width of strip I for $T\le 0$
\begin{eqnarray}\label{app a: L I}
L_{\text{I}}=\int_0^{z_{\max}} \frac{2\Big(\sqrt{h\left(z_{\max }\right)}-\lambda  h(z)\Big)\  dz}{h(z) \sqrt{H(z, z_{\max })-h(z_{\max })}},
\end{eqnarray}
where $z_{\max}=(1-T^2/4)^{1/3}$ from (\ref{app a: turning points}).  We stress that the above formula works for both the cases of the blue curve and the red curve of Fig. \ref{holo strip}. In particular, the integrand $2S'(z)\sim 2\Big(\sqrt{h\left(z_{\max }\right)}-\lambda  h(z)\Big)$ flips signs at turning point $z_c$ (\ref{app a: turning points}), which is the expected feature for the red curve. It suggests the integrand cannot be chosen as the absolute value $2|S'(z)|\sim 2\Big|\sqrt{h\left(z_{\max }\right)}-\lambda  h(z)\Big|$, which would miss the case of the red curve.   

\begin{figure}[t]
\centering
\includegraphics[width=6.8cm]{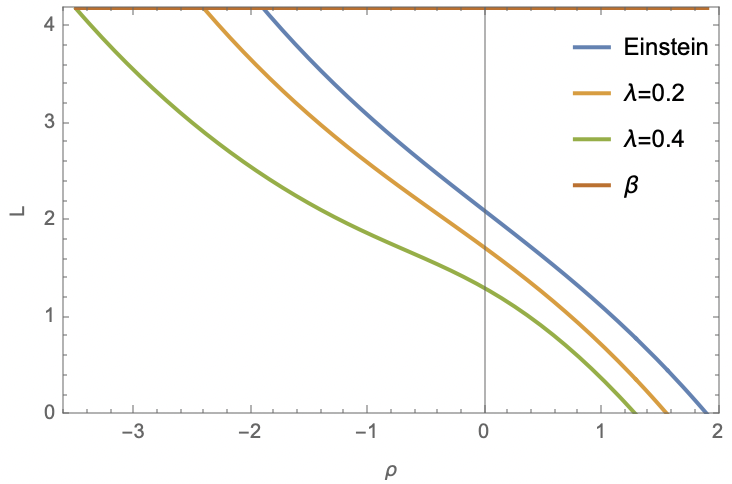} \includegraphics[width=6.8cm]{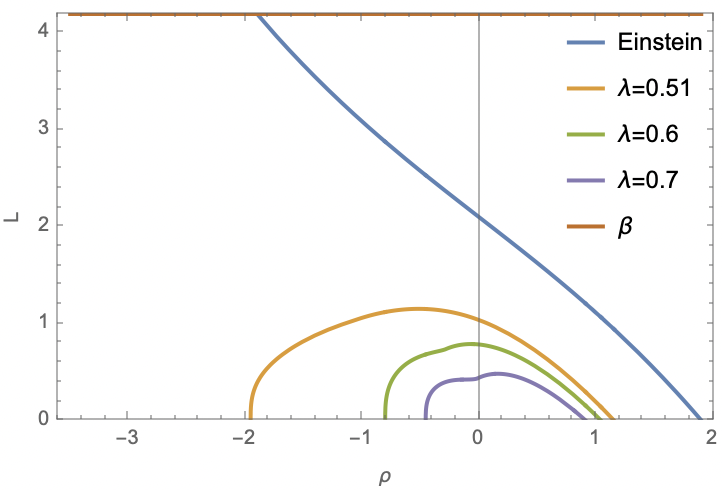}
\caption{Strip widths in normal phase with $0\le \lambda< 1/2$ (top) and in singular phase with $1/2< \lambda\le 1$ (bottom).  Note that $L$ can be larger than $\beta$ for negative enough $\rho$ in the normal phase. We show only range of $L\le \beta$ for simplicity. On the other hand, $L$ is always smaller than $\beta$ in the singular phase. $0\le L$ imposes a lower bound of $\rho$ in the singular phase, and a upper bound of $\rho$ in both phases. }
\label{strip width L1}
\end{figure}

Recall that strip II is the complement of strip I and $(T, \lambda)$ of NBC (\ref{app a: NBC}) flip signs when crossing the brane, we get the width of strip II for $T\ge 0$
\begin{eqnarray}\label{app a: L II}
L_{\text{II}}=\beta - L_{\text{I}}\Big(T\to -T, \lambda\to -\lambda\Big),
\end{eqnarray}
where, as we discuss above, the angle period $\beta=4\pi z_h/3$ is meaningful only for $T\ge 0$. In total, we have 
\begin{equation}\label{app a: L}
L=\begin{cases}
\ L_{\text{I}},&\
\text{for} \ T\le 0,\\
\  L_{\text{II}},&\
\text{for} \ T\ge 0.
\end{cases}
\end{equation}
One can check that $L$ is continuous at $T=0$, which is a test of our results. See Fig. \ref{strip width L1} for example. As mentioned in the main text, the constraint $0\le L$ sets an upper bound of $\lambda$. It is clear from (\ref{app a: L I}) that a large $\lambda$ would give a negative $L$. We numerically get the range $0\le \lambda\le 1$ for $L\ge 0$. As shown in Fig. \ref{strip width L1}, the case $0 \le \lambda \le 1/2$ can continuously transform into that of Einstein gravity, while the case $1/2 < \lambda \le 1$ cannot. We name them normal and singular phases, respectively. $\lambda=1/2$ is the phase-transition point.

Now we make some analytical discussions. 
Performing coordinate transformation $z=z_{\text{max}} y$ and expanding (\ref{app a: L I}) around $x=\text{sech}^2(\rho)\to 0$, we derive perturbatively for normal phase
\begin{eqnarray}\label{app a: L I 2}
&&L_{\text{I}}=\int_0^{1} dy \frac{2 \sqrt[3]{1-2 \lambda }}{\sqrt[6]{x} \sqrt{1-y^3}}+(1-2\lambda)\ O\left(\frac{x}{(1-2 \lambda )^2}\right)^{\frac{5}{6}} \nonumber\\
&=& \frac{2 \sqrt{\pi } \Gamma \left(\frac{4}{3}\right)}{\Gamma \left(\frac{5}{6}\right)}\Big(\frac{(1-2 \lambda )^2}{x}\Big)^{\frac{1}{6}}+(1-2\lambda)\ O\left(\frac{x}{(1-2 \lambda )^2}\right)^{\frac{5}{6}} \nonumber\\
\end{eqnarray}
The norm of the displacement operator becomes for $\rho<0$
\begin{eqnarray}\label{app a: CD}
C_D=\frac{32}{\frac{2 \pi  \lambda  \sqrt{x}}{1-2 \lambda  \sqrt{1-x}}+\pi  \cot ^{-1}\left(\sqrt{\frac{1}{x}-1}\right)}. 
\end{eqnarray}
Taking the limit $x\to 0$ ($\rho\to -\infty$) with $\lambda<1/2$, we obtain a universal limit
\begin{eqnarray}\label{app a: limit}
\lim_{\rho\to -\infty} (\frac{-\kappa_1}{C_D})=\lim_{x\to 0} (\frac{-L_{\text{I}}^3}{C_D})=-\frac{\pi ^{5/2} \Gamma \left(\frac{1}{3}\right)^3}{108 \Gamma \left(\frac{5}{6}\right)^3},
\end{eqnarray}
in normal phase. Note that $L_{\text{I}}$ (\ref{app a: L I 2}) goes to infinity for $x\to 0$. Recovering $z_h$ in $h(z)=1-z^3/z_h^3$, we can keep the strip width $L\to z_h L $ finite in the limit $x\to 0$ and $z_h\to 0$. Let us go on to discuss the case at the phase transition point $\lambda \to 1/2$. For our purpose, we consider the limit $x\to 0$, $\lambda\to 1/2$ with $x/(1-2\lambda)^2$ finite. In such limit, (\ref{app a: L I 2}) becomes
\begin{eqnarray}\label{app a: L I new}
L_{\text{I}}\to \frac{2 \sqrt{\pi } \Gamma \left(\frac{4}{3}\right)}{\Gamma \left(\frac{5}{6}\right)}\Big(\frac{(1-2 \lambda )^2}{x}\Big)^{\frac{1}{6}}. 
\end{eqnarray}
Interestingly, the second term of (\ref{app a: L I 2}) vanishes for $\lambda\to 1/2$. It is also the case for higher-order expansions. From (\ref{app a: L I new}) and $L_{\text{I}}=\beta=4\pi/3$, we solve 
\begin{eqnarray}\label{app a: proof 1}
x=\frac{729  \Gamma \left(\frac{4}{3}\right)^6}{64 \pi ^3 \Gamma \left(\frac{5}{6}\right)^6}(1-2 \lambda )^2+O(1-2\lambda)^3.
\end{eqnarray}
Substituting (\ref{app a: proof 1}) into (\ref{app a: CD}), we get
\begin{eqnarray}\label{app a: proof 2}
C_D=\frac{256 \sqrt{\pi } \Gamma \left(\frac{5}{6}\right)^3}{27 \Gamma \left(\frac{4}{3}\right)^3}+O(\lambda -\frac{1}{2}).
\end{eqnarray}
Finally, we obtain
\begin{eqnarray}\label{app a: proof 3}
\lim_{\lambda\to 1/2, L=\beta}(\frac{-\kappa_1}{C_D})=\lim_{\lambda\to 1/2}(\frac{-\beta^3}{C_D})=-\frac{\pi ^{5/2} \Gamma \left(\frac{1}{3}\right)^3}{108 \Gamma \left(\frac{5}{6}\right)^3},
\end{eqnarray}
which reproduces the universal limit (\ref{app a: limit}). Interestingly, $L=\beta$ instead of $L/\beta \to \infty$ at the phase transition point $\lambda \to 1/2$.



\begin{thebibliography}{00}


\bibitem{Casimir:1948dh}
H.~B.~G.~Casimir,
Indag. Math. \textbf{10}, no.4, 261-263 (1948)

\bibitem{Plunien:1986ca}
G.~Plunien, B.~Muller and W.~Greiner,
Phys. Rept. \textbf{134}, 87-193 (1986)

\bibitem{Bordag:2001qi}
M.~Bordag, U.~Mohideen and V.~M.~Mostepanenko,
Phys. Rept. \textbf{353}, 1-205 (2001)

\bibitem{Milton:2004ya}
K.~A.~Milton,
J. Phys. A \textbf{37}, R209 (2004)

\bibitem{Bordag:2009zz}
M.~Bordag, G.~L.~Klimchitskaya, U.~Mohideen and V.~M.~Mostepanenko,
Int. Ser. Monogr. Phys. \textbf{145}, 1-768 (2009)
Oxford University Press, 2009


\bibitem{Wang:2016och}
S.~Wang, Y.~Wang and M.~Li,
Phys. Rept. \textbf{696}, 1-57 (2017)

\bibitem{Vepstas:1984sw}
L.~Vepstas, A.~D.~Jackson and A.~S.~Goldhaber,
Phys. Lett. B \textbf{140}, 280-284 (1984)


\bibitem{Morris:1988tu}
M.~S.~Morris, K.~S.~Thorne and U.~Yurtsever,
Phys. Rev. Lett. \textbf{61}, 1446-1449 (1988)

\bibitem{Maldacena:2020sxe}
J.~Maldacena and A.~Milekhin,
Phys. Rev. D \textbf{103}, no.6, 066007 (2021)

\bibitem{Mohideen:1998iz}
U.~Mohideen and A.~Roy,
Phys. Rev. Lett. \textbf{81}, 4549-4552 (1998)

\bibitem{Bressi:2002fr}
G.~Bressi, G.~Carugno, R.~Onofrio and G.~Ruoso,
Phys. Rev. Lett. \textbf{88}, 041804 (2002)

\bibitem{Klimchitskaya:2009cw}
G.~L.~Klimchitskaya, U.~Mohideen and V.~M.~Mostepanenko,
Rev. Mod. Phys. \textbf{81}, 1827-1885 (2009)

\bibitem{Maldacena:1997re}
  J.~M.~Maldacena,
  Int.\ J.\ Theor.\ Phys.\  {\bf 38}, 1113 (1999)
  [Adv.\ Theor.\ Math.\ Phys.\  {\bf 2}, 231 (1998)]
  

\bibitem{Kovtun:2004de}
P.~Kovtun, D.~T.~Son and A.~O.~Starinets,
Phys. Rev. Lett. \textbf{94}, 111601 (2005)

\bibitem{Policastro:2001yc}
G.~Policastro, D.~T.~Son and A.~O.~Starinets,
Phys. Rev. Lett. \textbf{87}, 081601 (2001)

\bibitem{Brigante:2007nu}
M.~Brigante, H.~Liu, R.~C.~Myers, S.~Shenker and S.~Yaida,
Phys. Rev. D \textbf{77}, 126006 (2008)

\bibitem{Kats:2007mq}
Y.~Kats and P.~Petrov,
JHEP \textbf{01}, 044 (2009)


\bibitem{Brigante:2008gz}
M.~Brigante, H.~Liu, R.~C.~Myers, S.~Shenker and S.~Yaida,
Phys. Rev. Lett. \textbf{100}, 191601 (2008)

\bibitem{Miao:2024ddp}
R.~X.~Miao,
JHEP \textbf{06}, 084 (2024)



\bibitem{Jensen:2015swa}
K.~Jensen and A.~O'Bannon,
Phys. Rev. Lett. \textbf{116}, no.9, 091601 (2016)

\bibitem{Fujita:2011fp}
M.~Fujita, T.~Takayanagi and E.~Tonni,
JHEP \textbf{11}, 043 (2011)

\bibitem{Billo:2016cpy}
M.~Bill\`o, V.~Gonc{c}alves, E.~Lauria and M.~Meineri,
JHEP \textbf{04}, 091 (2016)

\bibitem{Bellucci:2009hh}
S.~Bellucci and A.~A.~Saharian,
Phys. Rev. D \textbf{80}, 105003 (2009)

\bibitem{McAvity:1993ue}
D.~M.~McAvity and H.~Osborn,
Nucl. Phys. B \textbf{406}, 655-680 (1993)


\bibitem{Miao:2018dvm}
R.~X.~Miao,
JHEP \textbf{07}, 098 (2019)

\bibitem{Romeo:2000wt}
A.~Romeo and A.~A.~Saharian,
J. Phys. A \textbf{35}, 1297-1320 (2002)

\bibitem{Toldin:2021kun}
F.~P.~Toldin and M.~A.~Metlitski,
Phys. Rev. Lett. \textbf{128}, no.21, 215701 (2022)

\bibitem{Ising referee1}
F. Parisen Toldin, S. Dietrich, J.Stat.Mech. 2010 P11003

\bibitem{Ising referee2}
M. Hasenbusch, Phys.Rev.B 82 (2010) 104425 

\bibitem{Cardy:1990xm}
J.~L.~Cardy,
Phys. Rev. Lett. \textbf{65}, 1443-1445 (1990)

\bibitem{Takayanagi:2011zk}
  T.~Takayanagi,
  Phys.\ Rev.\ Lett.\  {\bf 107} (2011) 101602

\bibitem{Dvali:2000hr}
G.~R.~Dvali, G.~Gabadadze and M.~Porrati,
Phys. Lett. B \textbf{485}, 208-214 (2000)


\bibitem{Miao:2023mui}
R.~X.~Miao,
JHEP \textbf{06}, 043 (2024)




\bibitem{deHaro:2000vlm}
S.~de Haro, S.~N.~Solodukhin and K.~Skenderis,
Commun. Math. Phys. \textbf{217}, 595-622 (2001)

\bibitem{Miyaji:2021ktr}
M.~Miyaji, T.~Takayanagi and T.~Ugajin,
JHEP \textbf{06}, 023 (2021)

\bibitem{Tadashi}
We thank Takayanagi for valuable comments on negative brane tension and its irrelevance to conical singularity.

\bibitem{Dantchev:2022hvy}
D.~M.~Dantchev and S.~Dietrich,
Phys. Rept. \textbf{1005}, 1-130 (2023)



\bibitem{Deutsch:1978sc}
D.~Deutsch and P.~Candelas,
Phys. Rev. D \textbf{20}, 3063 (1979)

\bibitem{Miao:2025utb}
R.~X.~Miao,
JHEP \textbf{04}, 023 (2025)


\bibitem{Miao:2017aba} 
  R.~X.~Miao and C.~S.~Chu,
  JHEP {\bf 1803}, 046 (2018)
  
\bibitem{Bloete:1986qm}
H.~W.~J.~Bloete, J.~L.~Cardy and M.~P.~Nightingale,
Phys. Rev. Lett. \textbf{56}, 742-745 (1986)


\bibitem{Bachas:2006ti}
C.~P.~Bachas,
J. Phys. A \textbf{40}, 9089-9096 (2007)

\bibitem{Diatlyk:2024qpr}
O.~Diatlyk, H.~Khanchandani, F.~K.~Popov and Y.~Wang,
Phys. Rev. Lett. \textbf{133}, no.26, 261601 (2024)

\bibitem{Krech:Casimir Effect}  
  M. Krech, The Casimir Effect in Critical Systems (World Scientific, London, 1994).
  
 
  
\bibitem{Hofman:2008ar}
D.~M.~Hofman and J.~Maldacena,
JHEP \textbf{05}, 012 (2008)

\bibitem{Hofman:2016awc}
D.~M.~Hofman, D.~Li, D.~Meltzer, D.~Poland and F.~Rejon-Barrera,
JHEP \textbf{06}, 111 (2016)
  
\bibitem{Bueno:2023gey}
P.~Bueno, H.~Casini, O.~L.~Andino and J.~Moreno,
Phys. Rev. Lett. \textbf{131}, no.17, 171601 (2023)
  
\bibitem{Collier:2021ngi}
S.~Collier, D.~Mazac and Y.~Wang,
JHEP \textbf{02}, 019 (2023)


 

\end{thebibliography}
\end{document}